# Dual orthogonally-polarized lasing assisted by imaginary Fermi arcs in organic microcavities

Teng Long,[1] Jiahuan Ren,[2] Peng Li,[3,4] Feng Yun,[4,5] Guillaume Malpuech,[6] Dmitry Solnyshkov,[6,7,*] Hongbing Fu,[1,*] Feng Li,[4,5,*] Qing Liao[1,*]

[1] Beijing Key Laboratory for Optical Materials and Photonic Devices, Department of Chemistry, Capital Normal University, Beijing 100048, People's Republic of China
[2] College of Physics Science and Technology, Hebei University, Baoding 071002, China
[3] Center for Regenerative and Reconstructive Medicine, Med-X Institute, The First Affiliated Hospital of Xi'an Jiaotong University, Xi'an 710061, China
[4] Key Laboratory for Physical Electronics and Devices of the Ministry of Education & Shaanxi Key Lab of Information Photonic Technique, School of Electronic Science and Engineering, Faculty of Electronic and Information Engineering, Xi'an Jiaotong University, Xi'an 710049, China
[5] Solid-State Lighting Engineering Research Center, Xi'an Jiaotong University, Xi'an 710049, China
[6] Institut Pascal, PHOTON-N2, Université Clermont Auvergne, CNRS, Clermont INP, F-63000 Clermont-Ferrand, France
[7] Institut Universitaire de France (IUF), 75231 Paris, France

* E-mails: dmitry.solnyshkov@uca.fr, hbfu@cnu.edu.cn, felix831204@xjtu.edu.cn, liaoqing@cnu.edu.cn

## Abstract

The polarization control of micro/nano lasers is an important topic in nanophotonics. Up to now, the simultaneous generation of two distinguishable orthogonally-polarized lasing modes from a single organic microlaser remains a critical challenge. Here, we demonstrate simultaneously orthogonally-polarized dual lasing from a microcavity filled with an organic single crystal exhibiting selective strong coupling. We show that the non-Hermiticity due to polarization-dependent losses leads to the formation of real and imaginary Fermi arcs with exceptional points. Simultaneous orthogonally-polarized lasing becomes possible thanks to the eigenstate mixing by the photonic spin-orbit coupling at the imaginary Fermi arcs. Our work provides a novel way to develop linearly-polarized lasers and paves the way for the future fundamental research in topological photonics, non-Hermitian optics, and other fields.

## Introduction

Polarized emission, particularly linearly-polarized lasing emission, has attracted increasing attention for its potential applications in integrated optoelectronic circuits, such as optical information processing, data storage, holography, and color display technology[1-5]. Thanks to the excellent laser properties and flexible molecular tailorability of organic gain materials, organic solid-state lasers have made great progress in the fields of the multiple-wavelength emitting and wavelength switching[6-8]. Importantly, organic single crystals (OSCs) with anisotropic properties are considered to be a promising lasing media, as their intrinsically highly polarized emission could lead to linearly polarized lasing[9-11]. As well known, the organic lasers operating at a certain optical transition usually have a single polarization[12]. This is because that the excitons with different polarizations in anisotropic OSCs have different optical gain and loss properties, and thus the thresholds in orthogonal polarizations are very different. Therefore, obtaining the simultaneously orthogonally-polarized lasing from a single OSC remains a critical challenge. It is desirable for the future dual-laser applications[13,14] including dual-comb generation[15-19], allowing to catch up with inorganic materials for lasing[20,21], which use schemes based on the Zeeman effect in magnetic fields.

Recently, non-Hermitian physics in nonconservative systems has attracted extensive interest, in particular for its chiral dynamics associated with non-Hermiticity by means of engineering gain and loss in optical systems[22-24]. Non-Hermiticity breaks the time-reversal symmetry without the need for external magnetic fields, which is extremely favorable for applications. Thus, the exceptional points (EPs), at which both the eigenvalues and eigenvectors coalesce and the Hamiltonian becomes nondiagonalizable, have enabled to engineer one-way light transport, enhanced sensors, and microlasers[25-28]. Experimental works demonstrated that an interchange of the instantaneous eigenvectors happened when a parametric variation winds a loop around an EP[29]. For the case of lasers as a typical non-Hermitian system, the exchanged modes can correspond to different polarization states[30]. Of a particular interest are exciton-polaritons, hybrid quasiparticles stemming from strong coupling between confined photons and excitons. They exhibit inherent non-Hermiticity due to the photonic and excitonic losses. EPs have been observed in various inorganic polaritonic systems since 2015[31-35]. Recently, some of us have demonstrated experimentally[36] the presence of the EPs without any artificial lattices or strong magnetic fields in organic exciton-polariton microcavities at room temperature, and observed the associated divergence of the quantum metric. Thus, the methods of non-Hermitian physics could be helpful to achieve orthogonally-polarized lasing in OSC systems. However, there are no experimental investigations on this concept of orthogonally-polarized dual laser assisted by non-Hermiticity in OSC exciton-polariton systems yet.

The transverse-electric transverse-magnetic (TE-TM) splitting is an example of a photonic spin-orbit coupling[37,38] (actually, the most widespread). It is broadly used in topological photonics[39,40] and spin-optronics[41]. In non-Hermitian systems, it was already shown to lead to two opposite effects: it can favor real[33] or imaginary[35] Fermi arcs, depending on the origins of the real and imaginary parts of the non-Hermitian effective field. Imaginary Fermi arcs are of the strongest interest for dual lasing, since they provide two modes with close eigenfrequencies and strongly modified eigenstates.

In this work, we present a demonstration of simultaneously orthogonally-polarized dual lasing from an OSCs-filled microcavity. The non-Hermitian effective field combining the TE-TM splitting and polarization-dependent lifetime mixes the eigenstates along the imaginary Fermi arcs in such a way that H-polarized excitons ultimately emit into both H- and V-polarized photonic modes. Our work provides a novel way to develop dual linearly polarized lasers and paves the way for the future of non-Hermitian optics, topological photonics, and other fields.

## Results

### Experimental system and its model

To demonstrate dual orthogonally-polarized lasing, we fabricate an optical microcavity, in which OSCs are sandwiched between two silver reflectors with the thickness of 100 nm (reflectivity ~ 99%) and 35 nm (reflectivity ~ 50%), as sketched in Fig. 1a. In order to prevent fluorescence quenching caused by direct contact between metallic layers and OSCs, 20-nm-thick $SiO_2$ thin-films were evaporated as space-layers between them. The active layer is a single-crystalline microbelts of an organic molecule, 1,4-dimethoxy-2,5-di(2,2',5',2''-terthiophenestyryl)benzene (TTPSB), whose structure is shown in Supplementary Materials (Scheme S1). The details on molecular synthesis, the fabrication of microcrystals and microcavity, and their characterization are depicted in Materials and Methods of Supplementary Materials. Here, the typical microbelt has the length (oriented along X direction) of about around 80 μm, the width (Y direction) of around 30 μm, and the thickness of about 1~1.5 μm (Supplementary Fig. S1). The crystal characterizations show the TTPSB molecules adopt a typical herringbone packing arrangement in the microbelts (Fig. S1f in Supplementary Materials), which brings about a significantly anisotropic molecular stacking density and strongly polarization-dependent excitonic absorption (as shown in Fig. S2 in the Supplementary Materials).

To understand the behavior of the system, we first construct and analyze a non-Hermitian effective Hamiltonian describing the strong coupling of two exciton modes and two photonic modes, as well as their linewidths, and the photonic spin-orbit

coupling. This Hamiltonian, written in the linear polarization basis, reads:

$$H_0 = \begin{pmatrix} E_{P,H} + \frac{\hbar^2 k^2}{2m_H} + \beta(k_x^2 - k_y^2) - i\Gamma_P & -2i\beta k_x k_y & V_H & 0 \\ 2i\beta k_x k_y & E_{P,V} + \frac{\hbar^2 k^2}{2m_V} - \beta(k_x^2 - k_y^2) - i\Gamma_P & 0 & V_V \\ 0 & 0 & E_{X,H} - i\Gamma_X & 0 \\ 0 & 0 & 0 & E_{X,V} - i\Gamma_X \end{pmatrix}$$

where $E_{P,H}$, $E_{P,V}$, $E_{X,H}$, $E_{X,V}$ are the bare photon and exciton energies in horizontal and vertical polarizations, respectively (at $k=0$); $\beta$ is the TE-TM splitting strength, $m_{H,V}$ are the cavity photon masses (approximately equal, due to the large mode numbers for both modes), $\Gamma_P$ is the cavity photon linewidth due to the finite lifetime, $\Gamma_X$ is the non-radiative linewidth of the excitons, and $2V_{H,V}$ are the Rabi splittings for the two polarizations. According to the absorption measurements presented in the Supplementary Materials (Fig. S2), the positions $E_{X,H}$, $E_{X,V}$ and the strengths $2V_{H,V}$ of the excitonic resonances in the two orthogonal polarizations are strongly different.

For two particular photonic modes *of the same parity* but with different mode numbers, a pair of lowest-energy eigenvalues of the Hamiltonian H$_0$ can look as shown in Fig. 1b. In V-polarization, the exciton resonance is high in energy and the Rabi splitting is small, so the V-polarized photonic mode (PM) is essentially uncoupled. The H-polarized exciton is much closer and stronger, so we can expect the formation of an exciton-polariton mode (EPM) with a much larger effective mass than the PM. These modes can have very different linewidths: the PM conserves the bare photon linewidth $\Gamma_P$, whereas the EPM presents a reduced linewidth $\Gamma_{EPM} = x\Gamma_X + (1-x)\Gamma_P$, where $x$ is the exciton fraction of the EPM. This pair of modes is therefore an essentially non-Hermitian system capable of hosting EPs on Fermi arcs. A similar configuration, but without strong coupling, has recently allowed to observe not only the presence, but also the motion and even the annihilation of EPs[35]. The orientation of Fermi arcs is determined by the TE-TM spin-orbit coupling (real effective field): along X and Y directions, it coincides with the imaginary effective field due to the polarization lifetime, and the PM and EPM branches cross. This is a real Fermi arc. Along the X-Y direction, the TE-TM splitting is orthogonal to the imaginary effective field, and the branches can anticross, showing an imaginary Fermi arc.

This is confirmed by the experimental measurements. We begin by studying the unpolarized reflectivity of the OSC-filled cavity, where the reflectivity is plotted as a function of wavelength (or energy) and angle (or momentum $k_x$), measured by angle-resolved micro-spectroscopy at room temperature (Schemes S2 and S3 in Supplementary Materials). Two sets of particular eigenmodes with different masses and very different linewidths, marked respectively by black and yellow lines, are observed. They cross in the X direction (Fig. 1c) and anticross in the X-Y direction (Fig. 1d). Additional results and polarization-resolved and orientation- dependent images confirming the H- and V-polarized nature of the PMs and EPMs are provided in Fig. S3 and Fig.S5 in Supplementary Materials.

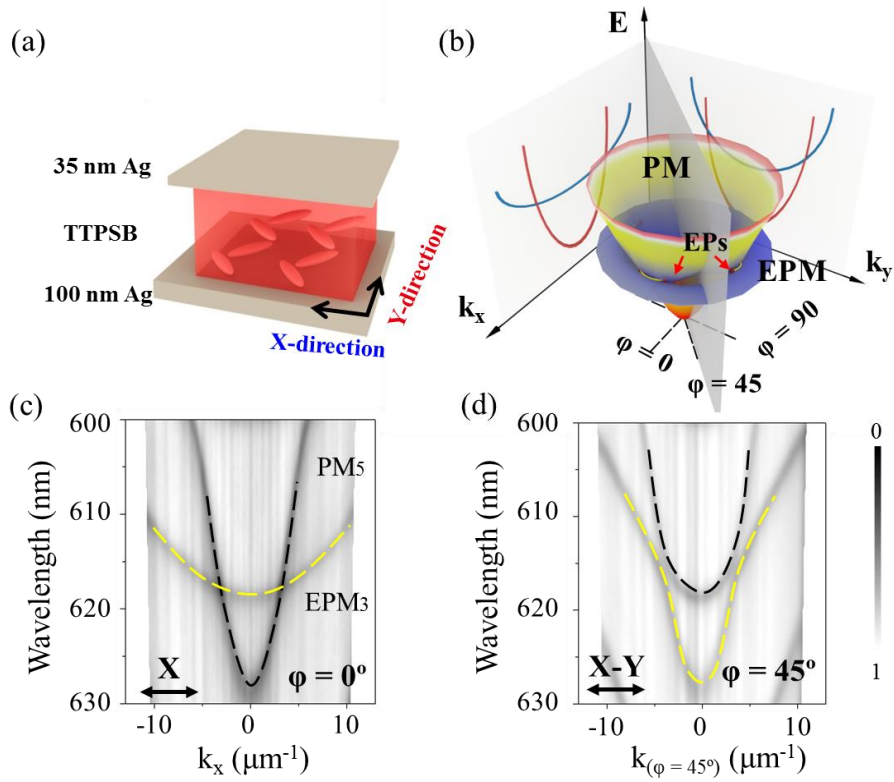

**Figure 1. The TTPSB cavity and its modes.** (a) Schematic diagram of TTPSB microcavity. (b) Schematic diagram of the distribution of the dispersions of both PM and EPM in k-space. (c,d) The angle-resolved reflectance spectroscopy of the crystal along the X direction (c) and X-Y direction (d). The numbers labeling the EPM and PM branches (i.e., EPM3 and PM5) are defined in Fig. S4 in Supplementary Materials.

**Experimental study of EPs and Fermi arcs**

In order to investigate the eigenvalues and the eigenstates of the OSC, we performed polarization-dependent angle-resolved reflectivity measurements. A linear polarizer, a half-wave plate and a quarter-wave plate were properly arranged in front of the spectrometer slit to obtain the polarization state of each pixel in the k-space. An energy spectrum is obtained in each of the six polarizations, i.e., linear polarizations (H, V, D, A) and circular polarizations (L, R) for each point of the reciprocal space. Fitting the modes of this spectrum with the pseudo-Voigt function allows determining the positions and the linewidths of the modes. The results of the extraction, proving the existence EPs and of the Fermi arcs connecting them, are shown in Fig. 2.

Fig. 2a shows the real part of the energy difference between the two modes as a function of the 2D in-plane wave vector $(k_x, k_y)$. The ellipse with relatively small values of the energy difference is composed of real Fermi arcs (magenta dashed lines, marked "RFA"), where the branches are degenerate, and imaginary Fermi arcs (black dash-dotted lines, marked "IFA"), where they split. Together with the imaginary part of the energy difference shown in Fig. 2c and exhibiting an opposite behavior (minima at imaginary Fermi arcs, maxima at real Fermi arcs), these results suggest the formation of exceptional points, where both real and imaginary parts of the energy difference are zero, up to the precision of the measurements. This is explicitly confirmed by Fig. 2e (dots with error bars), which shows the real and imaginary parts of the energy difference together, as a function of azimuthal angle along the ellipse formed by the Fermi arcs. The exceptional point is seen around 15.5°.

The results of the experiment are well reproduced by theoretical calculations (Figure 2(b, d)) based on the Hamiltonian $H_0$ (Eq. 1). The behavior of the real and imaginary parts of the energy along the ellipse is also well reproduced (solid lines in Fig. 2e). We have used the following parameters: $\beta = 0.4$ meV·µm² (extracted from the experiment), $m_H = 1.05 \times 10^{-5} m_0$, $m_V = 1.00 \times 10^{-5} m_0$, (both masses extracted from the experiment), $\hbar\Omega_H = 480$ meV, $\hbar\Omega_V = 240$ meV (twice smaller than in H, in agreement with the polarization-dependent absorption measurements shown in Fig. S2 in Supplementary Materials), $\Gamma_P = 10$ meV, $\Gamma_X = 0$ meV (the radiative broadening corresponding to the measured lifetime of 360 ps is negligible at this scale).

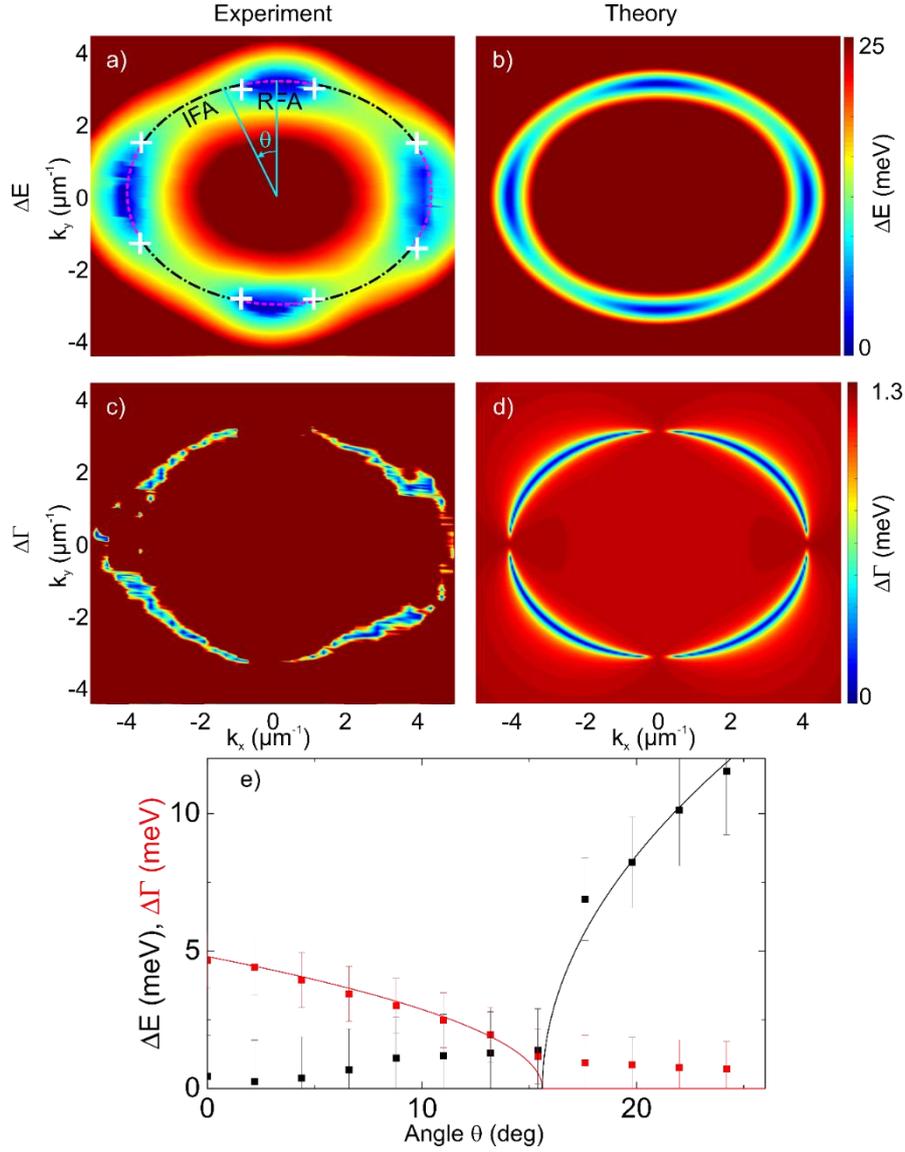

**Figure 2. Fermi arcs and exceptional points.** (a, b) Real part of the energy difference between the eigenmodes (experiment, theory), real (magenta dashed lines) and imaginary (black dash-dotted lines) Fermi arcs (marked RFA and IFA, respectively), white crosses mark the exceptional points, cyan lines mark the angle measurement for panel (e); (c, d) Imaginary part of the energy difference between the eigenmodes (experiment, theory); (e) Real (black) and imaginary (red) parts of the energy difference between the eigenmodes as a function of angle along the top Fermi arc marked RFA/IFA (dots with error bars marking the experimental uncertainty – experiment, solid lines – theory)

**Dual orthogonally-polarized polariton lasing**

Organic exciton-polaritons have attracted a large interest for their potential in polariton lasing and condensation at room temperature due to their large exciton binding energy and the Rabi splitting energy.[42,43] In order to study the behavior of polariton lasing, the 400-nm femtosecond laser was used to excite vertically our OSC-filled microcavity. The relaxation from the exciton reservoir to polariton states is assisted by vibrons.[44,45] Below threshold, strongest emission is thus observed from the EPMs, while there is very weak PL from the branches of PM, mainly focused near the EPs (Fig. S6 in Supplementary Materials).

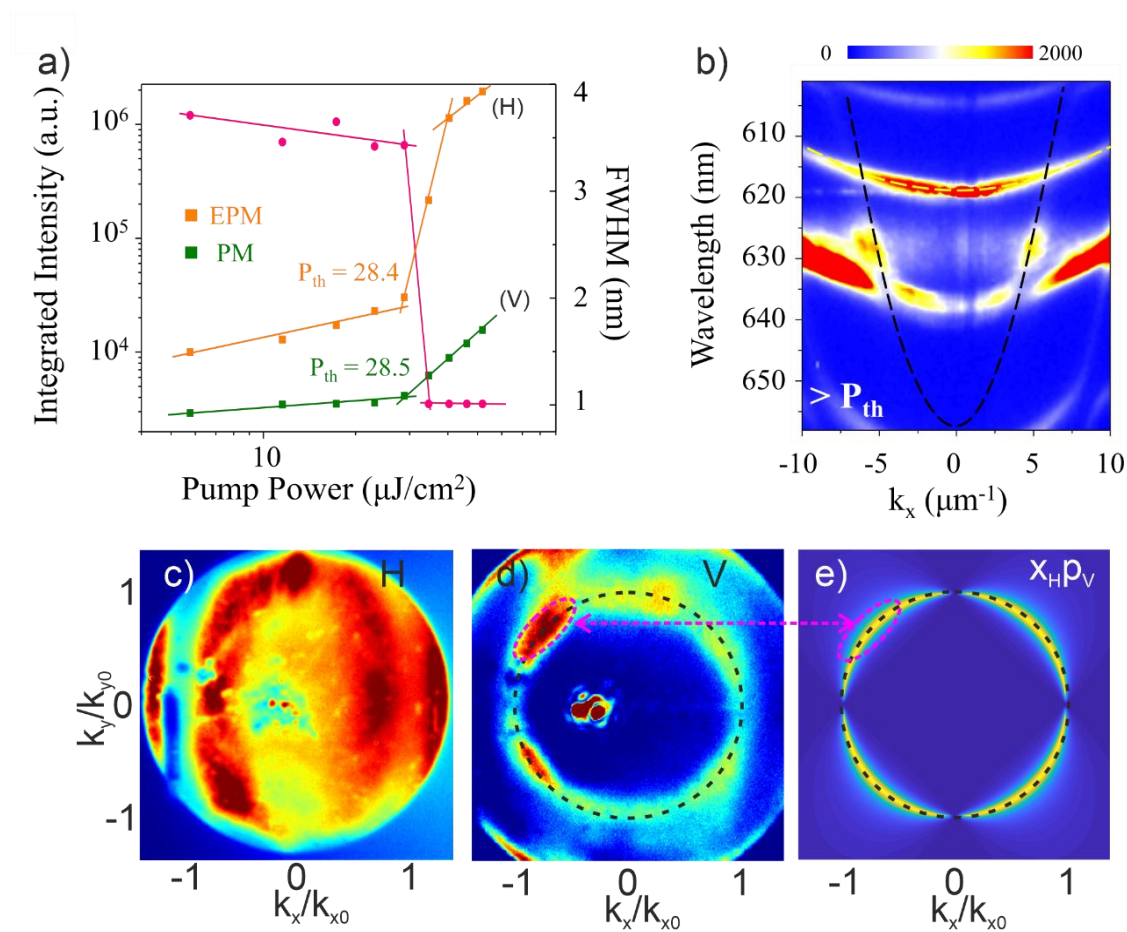

**Figure 3. Dual orthogonally-polarized polariton lasing.** (a) Integrated emission intensities of the H-polarized EPM3 (brown) and V-polarized PM (green) branches, and the FWHM (pink) of the EPM branch as a function of the pump fluence, demonstrating a lasing threshold and associated line narrowing. (b) Angle-resolved PL spectrum above the threshold, demonstrating a complex structure. (c) H-polarized emission distribution in reciprocal space: homogeneous emission at a constant energy determined by vibronic resonance, (d) V-polarized emission distribution in reciprocal space, demonstrating maxima at the imaginary Fermi arcs (The bright spot near the center is the pumping laser in V polarization which is not completely filtered.) (e) Product of V-photon and H-exciton fractions for the PM branch (theory), whose maxima determine the V-polarized emission in (d). Magenta arrow marks the product maximum responsible for V-lasing.

As the pump fluence is increased above the threshold, polariton condensation associated with the increase of PL intensity about 2 orders of magnitude is clearly seen along the H-polarized $EPM_3$ branch (Fig. 3a), measured at a slightly more negative detuning. The brown points in Fig. 3a show the integrated PL intensity of H-polarized $EPM_3$ emission as a function of the pump fluence, showing a typical threshold curve. The intensity dependence is separately fitted to power laws $x^p$ with $p = 0.6 \pm 0.07$ and $10.8 \pm 0.02$, respectively (straight lines). The threshold of the polariton lasing is $P_{th} = 28.4$ μJ·cm$^{-2}$ at the first intersection between the sublinear and superlinear regions. Meanwhile, the full width at half-maximum (FWHM) of the $EPM_3$ emission (magenta points and line) dramatically narrows from 3.65 nm below the threshold to 1.01 nm above the threshold.

Very surprisingly, lasing is also observed from the V-polarized branch $PM_5$ (green points in Fig. 3a), which is confirmed by the experimental measurements of the clear threshold ($P_{th} = 28.5$ μJ·cm$^{-2}$) via the change of the power laws (straight lines). A striking feature is the nearly equal values of lasing threshold between the EPM and the PM, despite the 100 times of difference of emission intensity between them below threshold. This undoubtedly suggests that the lasing emissions of EPM and PM originate from the same polariton condensation process. Figure 3b show the cut of the reciprocal space emission spectrum along $k_x$ for $k_y = 0$. The intensity distribution exhibits a complex structure, which nevertheless clearly demonstrates emission from both PM and EPM branches, confirming dual lasing. To understand the mechanisms of the dual lasing better, we study the full 2D reciprocal space below.

Fig. 3c shows the reciprocal space distribution of emission intensity in H-polarization (EPM) above threshold. The emission is mostly located at non-zero wave vectors, forming an elastic circle of approximately constant energy, determined by the vibronic resonance providing the fastest relaxation mechanism from the excitonic reservoir. The reciprocal space distribution of emission in V-polarization (PM) is much more interesting (Fig. 3d): its maxima are located at the position of the imaginary Fermi arcs, as compared with Fig. 2c,d. We note that the panels Fig. 3c-e are plotted in reduced coordinates defined by $k_{x,y}/k_{x0,y0}$ (with $k_{x0} = 4.1$ μm$^{-1}$ and $k_{y0} = 3.1$ μm$^{-1}$), where the set of Fermi arcs (real and imaginary) forms a perfect circle.

To explain the lasing from the imaginary Fermi arcs and understand its mechanisms, we calculate the product of the fractions of the H-exciton and V-photon in the PM branch, from which the lasing is observed, and plot it in Fig. 3e. The fact that the maxima of emission in Fig. 3d coincide with the maxima of fraction product in Fig. 3e demonstrates that in this region the PM branch is actually polaritonic, arising from an *indirect* strong coupling of H-exciton with V-photon mediated by the TE-TM spin-orbit coupling, in spite of the optical selection rules forbidding direct emission of H-excitons into V-photons. The lasing is therefore polaritonic, as for the H-polarized EPM branch, and this is why the thresholds are almost identical.

This band mixing is confirmed by the diagonal polarization degree extracted from the experiment below threshold (Fig. S8, dots with error bars, with examples of corresponding measurements in Fig. S7 in Supplementary Materials) along one quarter of the ellipse and compared with the theory (solid lines). The good agreement indicates a mixing of the eigenstates: in these directions they are not H and V, but a superposition of both, creating a possibility for dual lasing. It is also an additional confirmation of the validity of the model.

This original mechanism can be used not only for dual orthogonal-polarization lasing, but more generally to control the polarization and interaction properties of strongly and weakly coupled spin-optronic systems, both organic and inorganic. The spin-orbit coupling allows to modulate the selective strong coupling and to engineer the properties of optical systems.

**Conclusions**

In summary, we demonstrate simultaneously dual orthogonally-polarized lasing from an organic single crystal-filled microcavity at room temperature. We have shown the presence of exceptional points and Fermi arcs due to selective strong coupling. The lasing in both polarizations occurs via the polaritonic mechanism, with indirect strong coupling of H-excitons and V-photons ensured by the TE-TM spin-orbit coupling. Our work provides a novel way to develop dual-polarized lasers and paves the way for the future of topological photonics, non-Hermitian optics, and other fields.

**Acknowledgements**

This work was supported by the National Natural Science Foundation of China (Grant Nos. 22150005, 22090022, and 22275125), the National Key R&D Program of China (2022YFA1204402, 2018YFA0704805, and 2018YFA0704802), the Natural Science Foundation of Beijing, China (KZ202110028043), R&D Program of Beijing Municipal Education Commission (KM202210028016), Beijing Advanced Innovation Center for Imaging Theory and Technology. Shaanxi Key Science and Technology Innovation Team Project (2021TD-56), European Union's Horizon 2020 program, through a FET Open research and innovation action under the grant agreement No. 964770 (TopoLight), ANR Labex Ganex (ANR-11-LABX-0014), and by the ANR program

"Investissements d'Avenir" through the IDEX-ISITE initiative 16-IDEX-0001 (CAP 20-25).

F.L. acknowledges the Xiaomi Young Talents Program. The authors thank Dr. H.W. Yin from ideaoptics Inc. for the support on the angle-resolved spectroscopy measurements.

Supplementary Materials

# Dual orthogonally-polarized lasing assisted by imaginary Fermi arcs in organic microcavities

Teng Long,[1] Jiahuan Ren,[2] Peng Li,[3,4] Feng Yun,[4,5] Guillaume Malpuech,[6] Dmitry Solnyshkov,[6,7,]* Hongbing Fu,[1,]* Feng Li,[4,5]* Qing Liao[1,]*

[1]Beijing Key Laboratory for Optical Materials and Photonic Devices, Department of Chemistry, Capital Normal University, Beijing 100048, People's Republic of China

[2]College of Physics Science and Technology, Hebei University, Baoding 071002, China

[3]Center for Regenerative and Reconstructive Medicine, Med-X Institute, The First Affiliated Hospital of Xi'an Jiaotong University, Xi'an 710061, China

[4]Key Laboratory for Physical Electronics and Devices of the Ministry of Education & Shaanxi Key Lab of Information Photonic Technique, School of Electronic Science and Engineering, Faculty of Electronic and Information Engineering, Xi'an Jiaotong University, Xi'an 710049, China

[5]Solid-State Lighting Engineering Research Center, Xi'an Jiaotong University, Xi'an 710049, China

[6]Institut Pascal, PHOTON-N2, Université Clermont Auvergne, CNRS, Clermont INP, F-63000 Clermont-Ferrand, France

[7]Institut Universitaire de France (IUF), 75231 Paris, France

## MATERIALS AND METHODS

### 1. The structure of TTPSB molecule

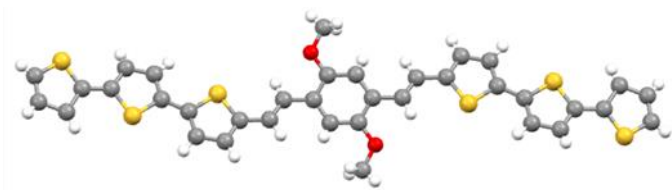

**Scheme S1.** The structure of the 1,4-dimethoxy-2,5-di(2,2',5',2''-ter-thiophenestyryl) benzene (TTPSB) molecule. The small balls represent carbon atoms (gray), hydrogen atoms (white), sulfur atoms (yellow), and oxygen atoms (red), respectively.

### 2. The preparation of TTPSB microbelts

In our experiment, TTPSB microbelts were fabricated using a facile physical vapor deposition (PVD) method. A quartz boat carrying 3 mg TTPSB was then placed in the center of a quartz tube which was inserted into a horizontal tube furnace. A continuous flow of cooling water inside the cover caps was used to achieve a temperature gradient over the entire length of the tube. To prevent oxidation of TTPSB, Ar was used as inert gas during the PVD process (flowrate: 15 sccm·min$^{-1}$). The pre-prepared hydrophobic substrates were placed on the downstream side of the argon flow for product collection and the furnace was heated to the sublimation temperature of TTPSB (at temperature region of ~ 320 °C), upon which it was physically deposited onto the pre-prepared hydrophobic substrates at temperature region of ~ 230 °C for 1 hours.

### 3. The preparation of TTPSB microcavity

Firstly, we use the metal vacuum deposition system (Amostrom Engineering 03493) to thermally evaporate silver film with the thickness of 85 ± 5 nm (reflectivity R ≥

99%) on the glass substrate, the root mean square roughness (Rq) of the silver film in the 5 μm × 5 μm area is 2.45 nm, a 20 ± 2 nm $SiO_2$ layer was deposited using vacuum electron beam evaporate on the silver film with $R_q$ of 2.31 nm, the deposited rates were both 0.2 Å/s and the base vacuum pressure is $3×10^{-6}$ Torr. This silver/$SiO_2$ film composite structure was placed as a substrate in a horizontal tube furnace for sample deposition. The TTPSB microbelts were uniformly dispersed on the silver/$SiO_2$ film substrate. Then 20 ± 2 nm $SiO_2$ and 35 ± 2 nm (R ≈ 50%) silver was fabricated to form the microcavity. The 20-nm $SiO_2$ layers are used to prevent the fluorescence quenching of TTPSB microbelts caused by directly contact of the metallic silver with the crystal.

## 4. The angle-resolved spectroscopy characterization

The angle-resolved spectroscopy was performed at room temperature by the Fourier imaging using a 100× objective lens of a NA 0.95, corresponding to a range of collection angle of ±60° (Scheme S2). An incident white light from a Halogen lamp with the wavelength range of 400-700 nm was focused on the area of the microcavity containing a TTPSB microbelt. The k-space or angular distribution of the reflected light was located at the back focal plane of the objective lens. Lenses L1-L4 formed a confocal imaging system together with the objective lens, by which the k-space light distribution was first imaged at the right focal plane of L2 through the lens group of L1 and L2, and then further imaged, through the lens group of L3 and L4, at the right focal plane of L4 on the entrance slit of a spectrometer equipped with a liquid-nitrogen-cooled CCD. The use of four lenses here provided flexibility for adjusting the magnification of the final image and efficient light collection. Tomography by scanning the image (laterally shifting L4) across the slit enabled obtaining spectrally resolved two-dimensional (2D) k-space images.

In order to investigate the polarization properties, we placed a linear polarizer, a half-wave plate and a quarter-wave plate in front of spectrometer to obtain the polarization state of each pixel of the k-space images in the horizontal-vertical (0° and 90°), diagonal (±45°) and circular (σ+ and σ− ) basis (S. Dufferwiel *et al.*, *Phys. Rev. Lett.* 2015, 115, 246401. & F. Manni *et al.*, *Nat. Commun.* 2013, 4, 2590.).

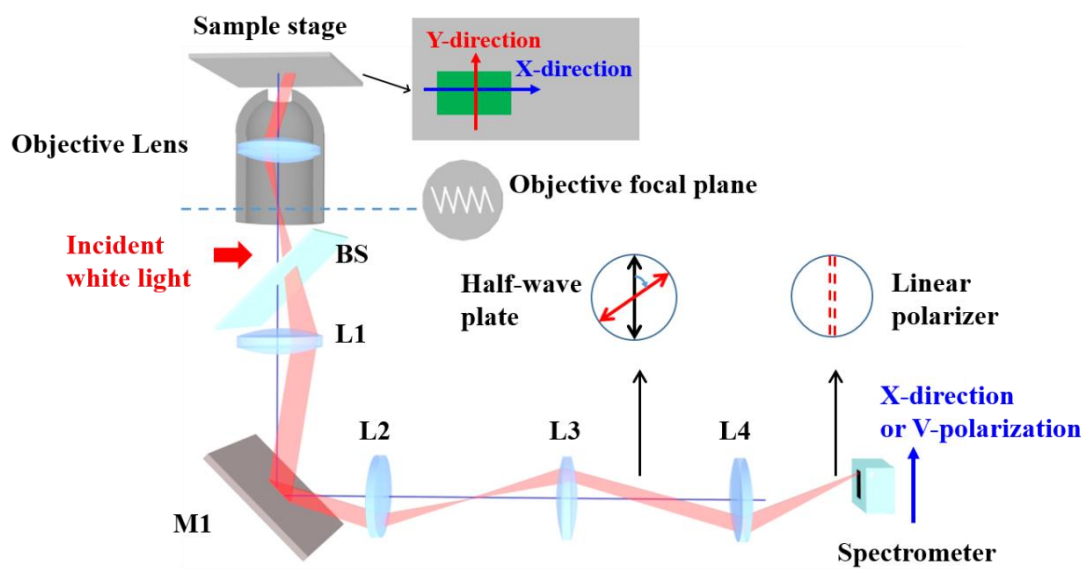

**Scheme S2.** Experimental setup allowing to obtain polarization-resolved complete state tomography. BS: beam splitter; L1-L4: lenses; M1: mirror. The red beam traces the optical path of the reflected light from the sample at a given angle.

The reflectivity measurement of our setup is shown below (Scheme S3). The reflectivity was also measured using a Halogen lamp with wavelength range of 400-700 nm. The light source was entered (green and white lines) and collected by using the same 100× microscope objective with a high numerical aperture (0.95 NA).

The measurement angle can achieve ±60°. The pink and white lines in the Figure below indicate the excited light path.

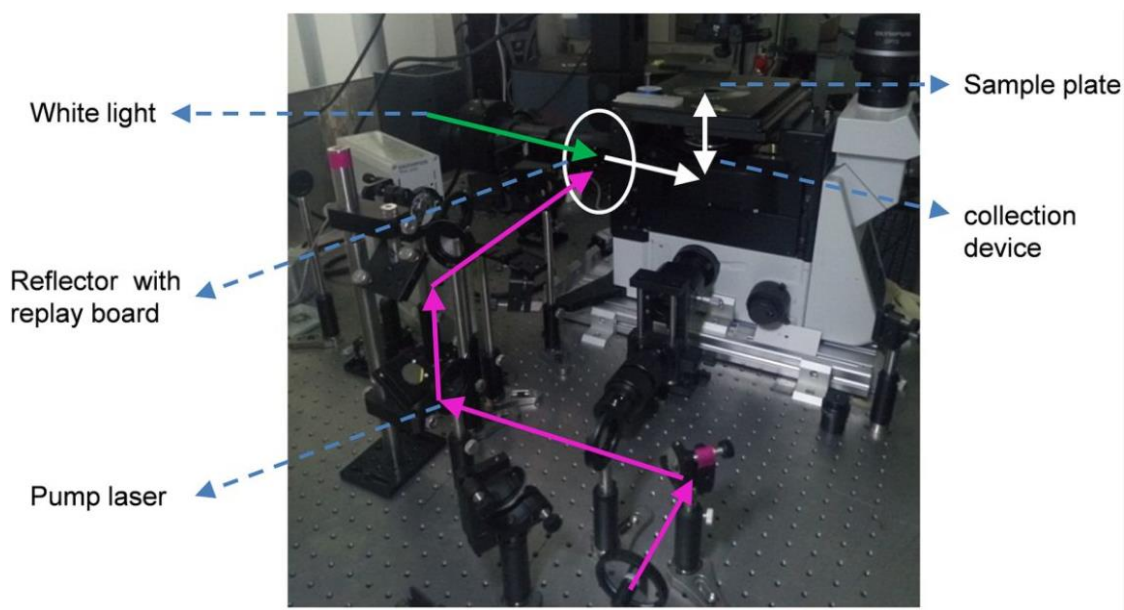

**Scheme S3.** The picture of our home-made experimental setup.

The reflectivity and photoluminescence spectroscopy was detected at room temperature in a home-made micro-area Fourier image which is presented to the spectrometer slit through four lenses, the schematic of the angle-resolved experimental setup is shown in Scheme S3. The reflected spectrum was collected by the spectrometer with a 300 lines/mm grating and a 400×1340 pixel liquid nitrogen cooled charge-coupled device (CCD). For off-resonant optical pumping (400 nm, pulse width 150 fs) from a 1 kHz Ti: sapphire regenerative amplifier and 40-μm spot diameter with a near Gaussian beam profile.

**S1 Crystal characterization**

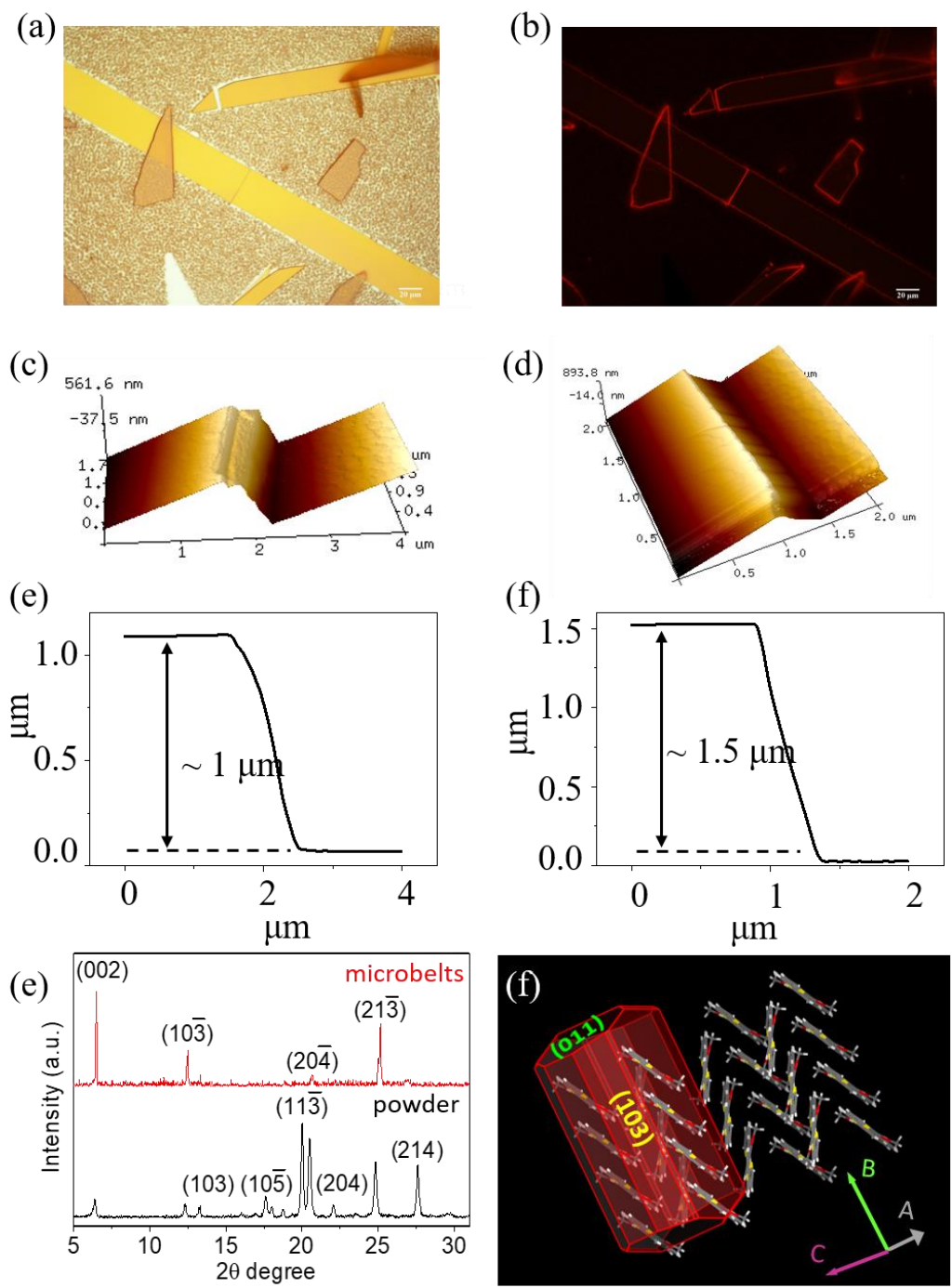

**Figure S1.** The bright (a) and dark (b) field images of TTPSB crystals. Atomic force microscopy (AFM) image of as-prepared TTPSB microbelts for sample 1 (c, e) and sample 2 (d, f). (e) XRD profiles of microbelts (red line) and powder (black line). (f) Simulated growth morphology of TTPSB molecules using Material Studio package.

**Table S1.** Crystal date and structure refinement for TTPSB.

| Space Group | P $2_1$/n |
|---|---|
| $a$ | 10.4006Å |
| $b$ | 5.54540Å |
| $c$ | 27.0574Å |
| $\alpha$ | 90.000º |
| $\beta$ | 93.409º |
| $\gamma$ | 90.000º |

**S2 Polarization absorption of pure crystal**

The anisotropy in the crystal is reflected in its absorption polarization response. In the direction parallel to the crystal Y-axis, the sample exhibits a very broad and intense absorption peak with a cutoff near 580 nm. In contrast, the absorption parallel to the X-axis direction exhibits three separate excitons peaks corresponding to 480 nm, 500 nm, and 540 nm, respectively.

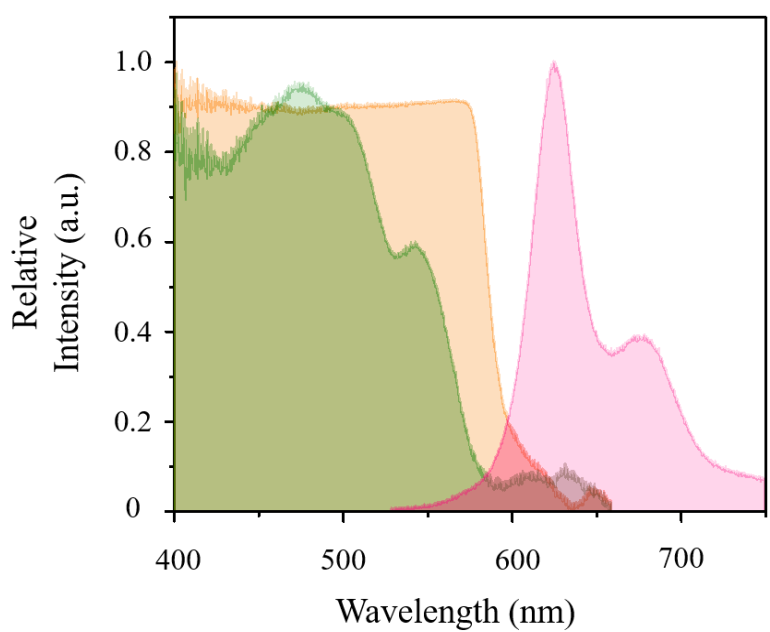

**Figure S2.** Polarization-dependent absorption (Y-direction: orange area; X-direction: green area) and emission (pink area) spectra of TTPSB single crystals.

**S3 Orthogonal linear polarization angle-resolved reflectance spectrum**

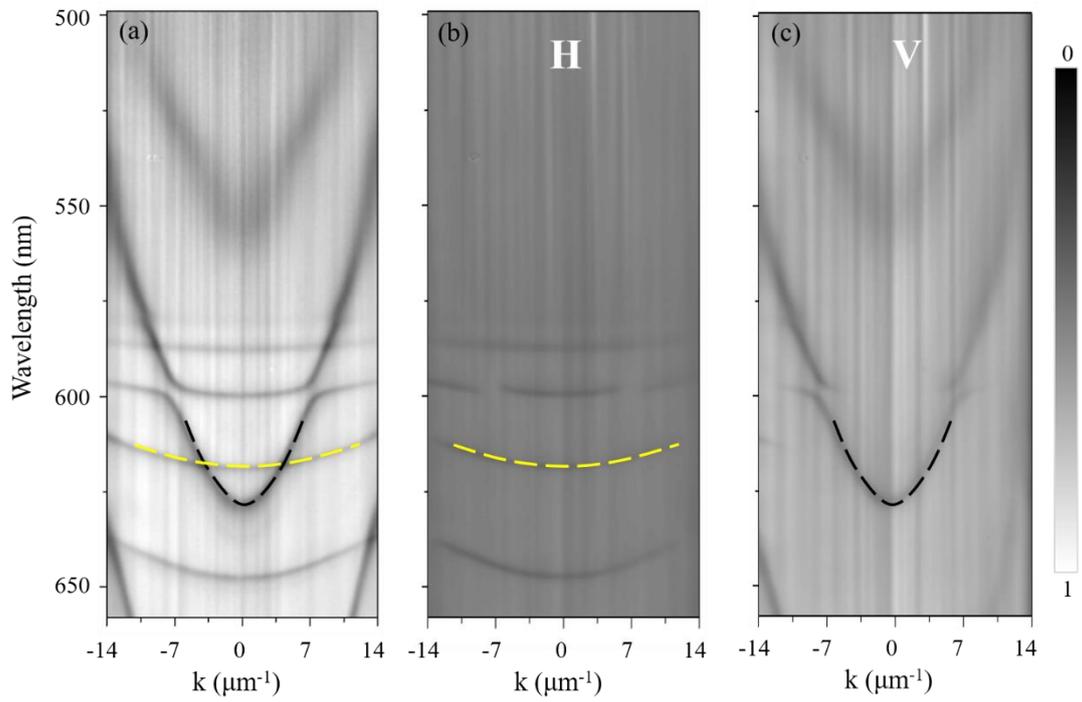

**Figure S3.** (a) Measured k-space angle-resolved reflectivity spectra of a selected microcavity at room temperature. The ARR in horizontal (H) polarization (b) and vertical (V) polarization (c) along X-direction of the single-crystal cavity.

## S4 Theoretical simulation of polaritons dispersion

The polariton dispersion in Fig. 1b was calculated by a coupled harmonic oscillator Hamiltonian (CHO) model (S. Kena-Cohen *et al.*, *Phys. Rev. Lett.* 2008, 101, 116401.). The 2×2 matrix in equation (1) below describes the CHO Hamiltonian:

$$\begin{pmatrix} E_{CMn}(\theta) & \Omega/2 \\ \Omega/2 & E_X \end{pmatrix} \begin{pmatrix} \alpha \\ \beta \end{pmatrix} = E \begin{pmatrix} \alpha \\ \beta \end{pmatrix} \quad (1)$$

Where $\theta$ represents the polar angle, $E_{CMn}(\theta)$ is the cavity photon energy of the $n^{th}$ cavity mode as a function of $\theta$, $E_X$ is the exciton 0–0 absorption energy of TTPSB microbelts at 2.11 eV (587 nm) and $\Omega$ (eV) denotes the coupling. The magnitudes $|\alpha|^2$ and $|\beta|^2$ correspond to the photonic and the excitonic fraction, respectively.

The cavity photon dispersion is given by:

$$E_{CMn}(\theta) = \sqrt{\left(E_c^2 \times \left(1 - \frac{\sin^2\theta}{n_{eff}^2}\right)^{-1}\right) - (n-1) \times l} \quad (2)$$

where $E_c$ represents the cavity modes energy at $\theta = 0°$, $E_{CM1}(\theta)$ represents the energy of the first cavity mode when n=1, $(n-1) \times l$ represents the energy difference from the first cavity mode. The effective refractive index ($n_{eff}$ = 1.8 and 3) is extracted from the fitting results. The theoretical fitting dispersion of the uncoupled cavity modes ($n_{eff}$ = 1.8, black solid line) and coupled cavity modes ($n_{eff}$ = 3, black dash line) is shown in Figure 1b. Diagonalization of this Hamiltonian yields the eigenvalues, $E_{\pm}(\theta)$, which represents the upper and lower polariton (UP and LP) in-plane dispersions (H. Deng *et al.*, *Rev. Mod. Phys.* 2010, 82, 1489-1537.),

$$E_{\pm}(\theta) = \frac{E_X + E_{CMn}(\theta)}{2} \pm \frac{1}{2}\sqrt{\left(E_X - E_{CMn}(\theta)\right)^2 + \hbar^2\Omega^2} \quad (3)$$

**Table S2. Coupled Harmonic Oscillator Model Fitting Results for EPM$_1$ to EPM$_4$.**

| Coupling mode | EPM1 | EPM2 | EPM3 | EPM4 |
|---|---|---|---|---|
| Rabi splitting (meV) | 180 | 480 | 480 | 480 |
| Detuning (eV) | 1.45 | 1.07 | 0.40 | 0.09 |
| $|\alpha_{ex}|^2$ | 0.99 | 0.95 | 0.82 | 0.59 |
| $|\beta_{ph}|^2$ | 0.01 | 0.05 | 0.18 | 0.41 |

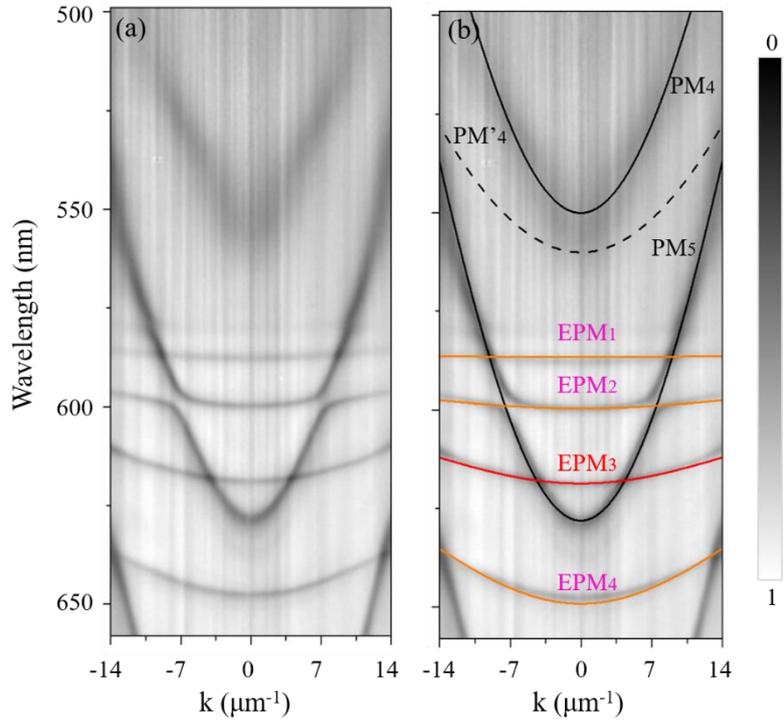

**Figure S4.** Angle-resolved reflectivity of the microcavity at room temperature in experiment (a) and theoretical simulated (b) results. EPM$_n$ denote the n-th exciton-polariton mode and PM$_n$ denote the uncoupled photonic mode (PM, black solid lines) respectively. PM'$_n$ denote the coupled photonic mode (PM', black dash lines), which derived from the theoretical simulation of the exciton polaritons.

**S5 AAR spectra at different placement angles**

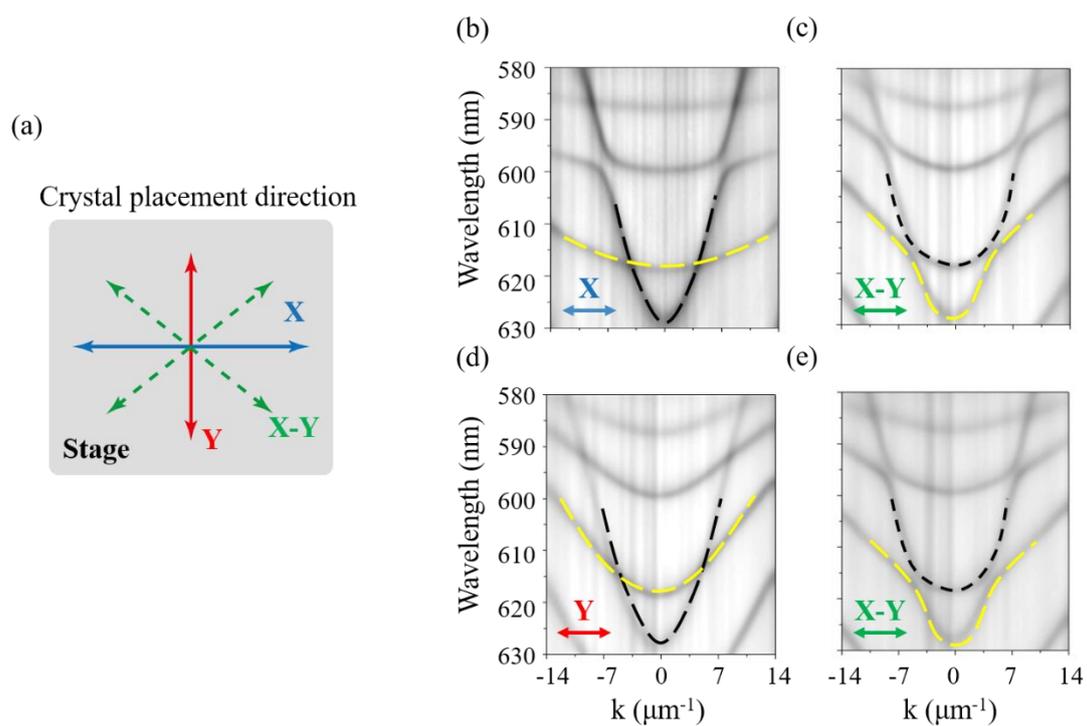

**Figure S5.** (a) Schematic diagram of the placement of crystals on the stage. (b)-(c) Angle-resolved reflectivity of the microcavity along the different placement of organic crystals.

## S6 Polarized emission below threshold

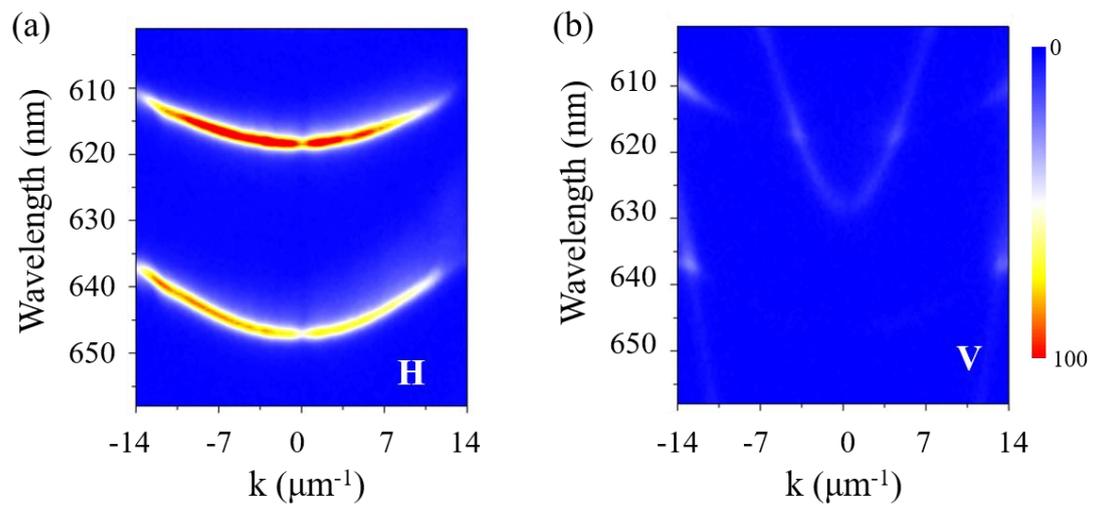

**Figure S6.** Emission on H (a) and V (b) polarization below threshold.

**S7 Quadrant reflectance maps of the 2D tomography in momentum space**

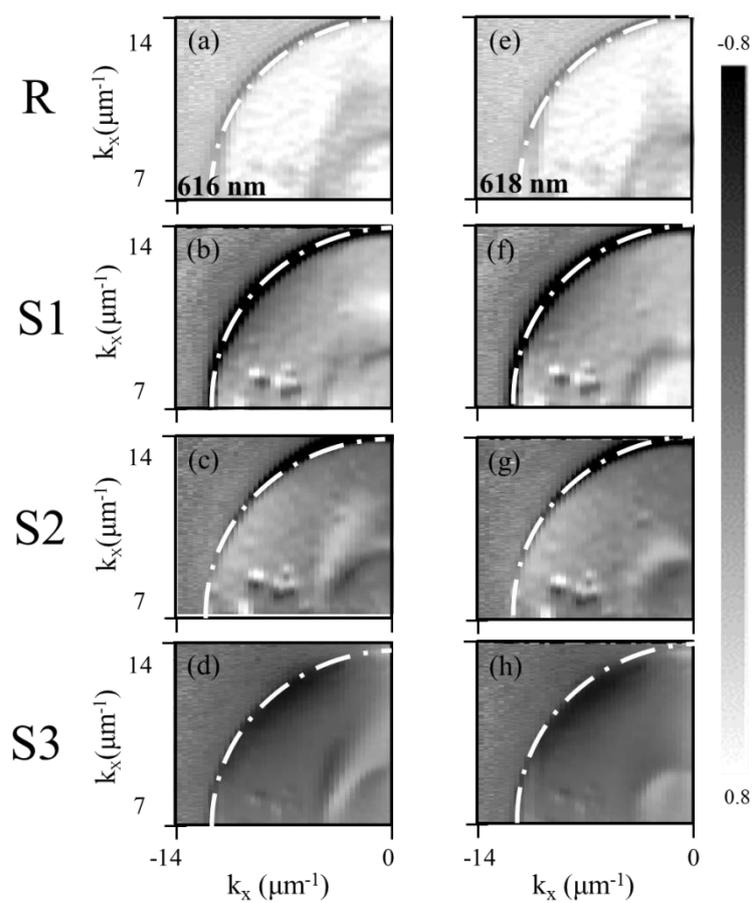

**Figure S7.** Quadrant reflectance maps of the 2D tomography at 616 nm (a) and 618 nm (e) in momentum space, respectively. Corresponding 2D wavevector maps of the Stokes vector components at 616 nm (b-d) and 618 nm (f-h). The white dotted line indicates the valid range.

**S8 The S2 Stokes vector component**

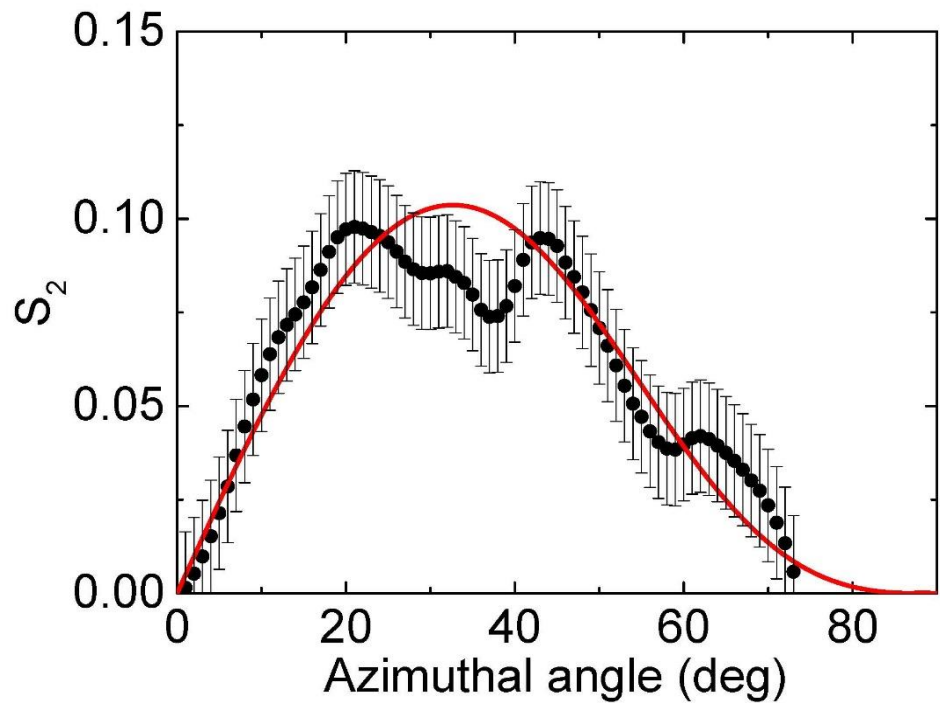

**Figure S8.** The S2 Stokes vector component (diagonal polarization degree) as a function of angle along the Fermi arc (dots with error bars marking the experimental uncertainty – experiment, solid lines – theory).